\documentclass[12pt,preprint]{aastex}

\shorttitle{}
\shortauthors{G\'omez de Castro, A.I.}

\begin{document}


\title{AK Sco, first detection of a highly disturbed atmosphere
in a pre-main sequence close binary\\}


\author{Ana I. G\'omez de Castro}
\affil{S.D. Astronomia y Geodesia, Fac. de CC Matem\'aticas,
Universidad Complutense de Madris, E-28040 Madrid, Spain}
\email{aig@mat.ucm.es}




\begin{abstract}
AK Sco is a unique source: a $\sim$10 Myrs old pre-main sequence
spectroscopic binary composed of two nearly equal F5 stars that at
periastron are separated by barely eleven stellar radii so, the
stellar magnetospheres fill the Roche lobe at periastron. The
orbit  is not yet circularized (e=0.47) and very strong tides are
expected. This makes of AK Sco, the ideal laboratory to study the
effect of gravitational tides  in the stellar magnetic field
building up during pre-main sequence (PMS) evolution. In this letter,
the detection of a highly disturbed  ($ \sigma  \simeq
100$~km~s$^{-1}$) and very dense atmosphere (n$_{\rm e} = 1.6
\times 10^{10}$~cm$^{-3}$) is reported. Significant line broadening 
blurs any signs of ion belts or bow shocks in the spectrum of
the atmospheric plasma. The radiative loses
cannot be accounted solely by the dissipation of energy from the
tidal wave propagating in the stellar atmosphere; neither by the
accreting material. The release of internal energy from the star
seems to be the most likely source of the plasma heating. This
is the first clear indication of a highly disturbed atmosphere
surrounding a pre-main sequence close binary.

\end{abstract}





\section{Introduction}

Pre-main sequence (PMS) stars are surrounded by very powerful
magnetospheres that extend 2-4 stellar radii to the limits of the
inner disk boundary. The magnetospheres are powered by the stellar
magnetic field, the accreting infalling material and the
shearing at the boundary between the magnetosphere and the inner
disk (see G\'omez de Castro 2009 for a recent review).
Magnetospheric heating processes during pre-main sequence
evolution are poorly studied, as well as the relative relevance of
the contributions mentioned above. This is partially caused by the
similar densities (10$^9-10^{11}$ cm$^{-3}$) and temperatures
(10$^{4.5}-10^{4.8}$K) of several distinct physical components in
the circumstellar environment (the disk-star shear layer, the
outflow and the accretion flow). Moreover, the high temperature of
the plasma makes thermal broadening very large ($\sim
70$km~s$^{-1}$) leaving ultraviolet high spectral resolution
monitoring as the only technique to resolve the various
contributions since variability time scales are expected to be
very different for the various components (see i.e. G\'omez de
Castro \& Verdugo 2007).

Close pre-main sequence binaries are ideal targets to unveil some
of the magnetospheric properties. Circumstellar accretion disks
are difficult to accommodate in these systems since the strong
gravitational tides break them into rings. The inner boundary of the
circumbinary disk is pushed outwards by the binary orbit thus the
magnetosphere of each individual star is not locked to the disk
rather, it is expected that a strong coupling between the gravitational tides and the stellar magnetic fields ought to be a major source of atmospheric/magnetospheric heating. The investigation of the
atmospheric properties of close pre-main sequence stars
provides important clues regarding the possible role of magnetospheric
dynamos in pre-main sequence evolution.

AK Sco is a unique target for this purpose; it is the only
pre-main sequence binary known to date with small eccentricity and
composed of two nearly equal stars that get as close as 11~R$_*$
at periastron passage. No signs of enhanced accretion have been
detected at periastron (Andersen et al. 1989, Alencar et al.
2003). Polarimetric observations have reported the largest
variations of percent polarization and position angle detected in
a pre-main sequence binary (Manset et al. 2005). Variations are
found to be periodical with period equal to the gravitational tide
period (half the orbital period). The main parameters of the
binary are summarized in Table~\ref{tbl-1}. In this letter, the
profiles of the (semiforbidden) intercombination lines of the
optically thin magnetospheric tracers: the CIII] at 1908\AA\ and
the Si III] at 1892\AA\ spectral lines, are analyzed. From the
analysis, it is concluded that there is an extended region  
of radius $\sim 1.3-4.9$R$_*$ around each star containing hot 
($\log T_{\rm e} = 4.8$) and highly perturbed gas that produces
a gaussian-like profile broadening with $\sigma \sim 100$~km~s$^{-1}$. Although this broadening is highly suprathermal it could be
sufalv\'enic if  AK~Sco's magnetic field is similar or higher than
that reported for other pre-main sequence stars (see i.e. Johns-Krull,
2007).

\section{Observations and measurements}

AK~Sco was observed with the Space Telescope Imaging Spectrograph (STIS)
on March 10, 2001. The observations were carried out with grating G230M to target the semiforbidden lines of Si~III] at 1892\AA\ and C~III] at 1908\AA . The spectra were obtained at phase 0.2814 according to Andersen et al (1989)
ephemeris (see Table~1). At this phase, the radial velocity of both components is the same thus the profiles represent the co-added contribution of both components in the same velocity space. As all indications are that the two components are equal in all respects, including spectral type, the contribution from both stars is expected to be the same.

AK~Sco displays strong and very broad Si~III] and C~III] profiles.
The observed line profiles have been converted to velocity
distribution using the Doppler shift formula. Data have been
processed with the Routine Science Data Pipeline (RSDP). The major
source of  inaccuracy in the calibration is the centering of the
target in the aperture that can account for as much as
2.8~km~s$^{-1}$.  The exposure duration was 1944~s that were
splitted into three subexposures; the average value for each pixel
and the deviation from this average are plotted in Fig.~1. A curve
fit to a single gaussian plus a constant is also shown. The
dispersion of the gaussians are $\sigma$(C~III])=109.5~km/s and
$\sigma$(Si~III])=106.8~km/s. The value of $\chi ^2/\nu$ is 3.6
and 1.3 for this best gaussian fit to the Si~III] and C~III]
lines, respectively. The instrumental profile of HST/STIS with the
grating G230M has a FWHM of 30~km/s and the stellar rotation
velocity is $v \sin i = 18.5$~km~s$^{-1}$, thus line
broadening\footnote{The FWHM of the profiles is
FWHM(Si~III])=251~km/s and FWHM(C~III])=258~km/s.} is more than
one order of magnitude larger than expected from the rotational
broadening of the atmospheric lines.

The lines fluxes are: F(Si~III]) = $(4.2 \pm 0.1)\times
10^{-13}$erg/s/cm$^2$ and F(C~III]) = $(2.9 \pm 0.1)\times
10^{-13}$erg/s/cm$^2$, after correction for extinction using
Valencic et al. 2004 UV law (R= A$_V$/E$_{B-V}$=4.3 from Manset et
al. 2005 and A$_V = 0.5 \pm 0.1$ mag from Alencar et al. 2003).
The lines ratio is calculated to be F(Si~III])/F(C~III]) = $1.4
\pm 0.2$. This value is similar to that derived from earlier (1982
and 1988) low resolution observations obtained with the
International Ultraviolet Explorer (IUE) (G\'omez de Castro \&
Franqueira 1997). From the IUE low dispersion spectra
a C~IV($\lambda \lambda1548,1550\AA)$ flux of $(1.6 \pm
0.1)\times 10^{-12}$erg/s/cm$^2$ and a ratio F(C~IV)/F(Si~III]) of
$1.9\pm 0.4$ was derived. There are only three low resolution observations
carried out with the IUE, two of them during the same cycle at
periastron (phase=0.123) and at apastron (phase=0.708).  No
significant variations in the C~III], Si~III] and C~IV
fluxes\footnote{There are however, variations by a factor of $\sim
2$ in the O~I ($\lambda 1304$\AA ) and C~II resonance multiplet at
1335\AA ; these lines trace cooler and lower density plasma.} are
found though the IUE fluxes are about a factor of $\simeq 2$
higher than that measured with the HST.

\section{Constraints from the lines flux}

Assuming collisional equilibrium, the F(Si~III])/F(C~III]) and
F(C~IV)/F(Si~III]) ratios indicate that the physical conditions in
this Hot Line Emission Region (HLER) are $\log n_e({\rm cm}^{-3})
= 10.2$ and $\log T_e({\rm K}) =4.8$. An optically thin
collisional plasma with these properties radiates $2.4\times
10^{-3}$ of the total flux in the C~III] line; thus the luminosity
of the HLER is 3.1$\times 10^{32}$erg/s or 0.08$L_{\odot}$
(0.04$L_{\odot}$ per star assuming the same contribution per
component); a distance, $d$, of 145~pc to AK~Sco has been assumed.
To account for this high luminosity an accretion rate\footnote{A
radius of $0.43$~AU (three times the semimajor axis of the orbit)
is estimated for the gap of the circumbinary accretion disk based
on numerical simulations of the evolution of binary systems
(Artymovicz \& Lubow, 1997).} of $\dot M_a = 0.9\times 10^{-7}$
M$_{\odot}$yr$^{-1}$ is required.  The viscous release of this
gravitational energy in the circumbinary disk would affect
significantly the infrared spectral energy distribution (SED)
however, good fits are obtained with passive disk models
suggesting that the accretion rate is not very much above $1\times
10^{-8}$~M$_{\odot}$yr$^{-1}$ (see i.e. Alencar et al. 2003 for a
discussion on the SED fit).

The characteristic size of the HLER, $R_e$, is given by,
$$
f \left[ \left(\frac {R_e}{R_*} \right)^3 -1\right] = \frac {3 F_l
d^2} {< j_l > R_*^3} =   0.4
$$
where $R_e$ is the radius of an equivalent homogeneous spherical
envelope around each star,  with $<j_l>$ representing the average emissivity in
the (Si~III]) lines formation region, which for this study, has been assumed 
to be of the order of $j_l = 2.38 \times
10^{-4}$erg~s$^{-1}$~cm$^{-3}$ using the atomic
parameters from the Chianti Atomic Data
Base\footnote{URL:www.damtp.cam.ac.uk/user/astro/chianti/}).  Here
$f$ represents the volume filling factor of the radiating plasma.  For $f
\sim 0.1$, $R_e$ reaches 1.7~R$_*$. This is an indication of the
HLER being significantly more extended than the atmosphere of a
cool main sequence star.

\section{Constraints from the line broadening}

Line broadening is highly suprathermal (thermal broadening at
$10^{4.8}$K generates a FWHM of 73~km/s, a factor of $\sim 3$
smaller than the observed). In principle, such a broad kinetic
velocity distribution of the emitting ions could be produced, as in 
the Sun, by resonances between Alfv\'en waves propagating in the atmosphere
and the giro-frequency of the ions, driving the Si~III] and C~III]
ions to the very high kinetic temperature of $T_{\rm kin} =
0.75\times 10^6$K (see i.e. Cranmer 2002).
However, the strength of this broad component must correlate with
the coronal X-ray surface flux produced by the bremsstrahlung
radiation of the high velocity electrons, as shown for
magnetically active stars by Wood et al. 1997. Though AK~Sco X-ray
surface flux, $F_X = 6.49 \times 10^5$~erg/s/cm$^2$ (derived from
Steltzer et al. 2006 based on CHANDRA observations) is similar to
the detected in class IV-V F5 stars (see Ayres et al. 1995), the
CIV flux exceeds by $\sim 4.5$ orders of magnitude the  flux predicted
by the active stars CIV-X~ray surface flux correlation. Thus, the
HLER radiation cannot be accounted for solely by this mechanism.

The high symmetry of the profiles is suggestive of them being
formed in a rotating ionized ring or belt such as the observed
around RW~Aur (G\'omez de Castro \& Verdugo 2003). However the
characteristics of AK~Sco orbit imposes severe constraints on the
possible location of the material with Keplerian-like orbits. On
the one hand, circumstellar material is forced to be between the
stellar corona ($\sim 2 R_*$) and the location of the Lagrange
point at periastron (5.5$R_*$). Plasma in such a hypothetical ring
systems surrounding each star would be orbiting with velocities in the
range $180 < V_{\rm ring} < 280$km~s$^{-1}$ and thus, could only
contribute to the line wings. On the other hand, the binary cleans
the inner region of the circumbinary disk to about 0.43~AU thus,
any ionized material in this boundary would be rotating at
Keplerian velocities $\sim 75$km~s$^{-1}$. As a consequence, the
profile broadening can neither  be accounted for  by rotating rings
around the stars nor by a circumbinary belt. Moreover the low
rotation velocity of the stars precludes that line broadening can
be accounted by an extended magnetosphere, rigid-body like
rotation as shown in Fig~2. Line profiles have been simulated for
various emissivity and density profiles for a corotating
magnetospheric shell and belt. The energy radiated in the lines
per unit volume, $\epsilon _{rad,l}$, at a given distance, $r$,
from the center of the star is given to first order as: $\epsilon
_{rad,l} = j(r)\times \rho (r) \propto r^{\alpha}$ where $j(r)$
represents the line emissivity and $\rho$ the density of emitting
ions. As shown in Fig~2, simple spherical and cylindrical
distributions (simulating a shell-like or a belt-like structure,
respectively) cannot account for the observed widths of the HLER
lines. The optimal fit is obtained by adding a {\it
Maxwellian velocity distribution} with variance from
$\sigma =$74~km/s for a corotating ring peaking at d=5.5$R_*$
($\chi ^2$ = 27.5 and $\chi ^2/\nu$ = 1.44) to
$\sigma =$88~km/s for a spherical envelope with $\epsilon _{rad}
\propto  r^{-2}$ ($\chi ^2$ = 19.5 and $\chi ^2/\nu$ = 1.2).
In all cases, the Maxwellian velocity field is highly supersonic and blurs
the possible kinematical signatures of rotating structures around
the star.

\section{Conclusions: on the source of the HLER heating}

AK~Sco has an UV excess that accounts by as much as 0.04~L$_{\odot}$
per component. This energy cannot be fed into the system solely by
the gravitational energy lost by the accretion flow channelled by the surrounding disk since the accretion rates required are about one order of magnitude larger than that derived from the infrared properties of the disk. Even if the contribution
of the fresh deuterium to the nuclear reactions is taken into account
(Siess et al 1997), the accretion rate is a factor of ten smaller than
that required to produce the observed luminosity\footnote{An estimate of the luminosity excess caused by accreting fresh deuterium can be made from the nuclear energy released per reaction is $Q_D=5.5$~MeV. The luminosity released is given by:
$
L_{D} = (1/2) (m_D/m) n_A \dot M_a Q_D
$
where $m_D/m$ is the mass fraction of deuterium in the accreting matter,
$n_A$ the Avogadro number and the factor of (1/2) is introduced to
account for the Deuterium mass number. Assuming a mass fraction deuterium
to hydrogen of $10^{-5}$, solar abundances and an accretion rate of
$1.0\times 10^{-8}$~M$_{\odot}$yr$^{-1}$, the luminosity excess will
be 0.004~L$_{\odot}$.}.  

The gravitational energy stored in the epicyclic motion of the
relative orbit is $\epsilon _o = \frac {1}{2} \frac {M_1 M_2}
{M_1 + M_2} e^2 \omega ^2 a^2 = 2.36 \times 10^{46}$erg where
$M_1$ and $M_2$ are the masses of the two components of the system,
$e$ the eccentricity, $a$ the semimajor axis and $\omega = \frac {2 \pi}
{P}$ with $P$ the period. If the gravitational energy were dissipated through the HLER radiation, it
would be damped in a time scale of: $ \tau _{tide} = \epsilon _o
/0.08L_{\odot} = 2.4 \times 10^6 {\rm yr}^{-1} $ which is rather close to the age of AK~Sco derived from the evolutionary tracks and the
system should be already circularized. Henceforth, an additional source
of energy needs to be hypothesized to power the HLER. This source
also, should be acting in other pre-main sequence stars and be partially  responsible of the observed UV excess.

Finally, notice that the magnetic diffusivity in the HLER, $\eta$, is very small (just 0.0045 cm$^2$/s) leading to very high Reynolds magnetic numbers
and magnetic flux freezing. This has two important consequences:
(1) the tidal deformation wave induces magnetic multipoles into
the stellar dipolar fields that can be responsible of the
unusually strong periodic polarimetric  variations of the system
and (2) the Alfv\'en waves induced by the tide will have very large 
damping lengths due to magnetic resistivity effects. The non-linear 
wave-wave interaction will lead to a turbulent cascade that will
end at a scale determined by the dominant damping mechanism.
For the HLER physical properties, this will be collisions between the ionized HLER material and the neutral infalling gas from the accretion flow
that, in turn, strongly depends on the ill determined relative densities
of these two components. If the observed broadening of the HLER lines is 
interpreted as a sign of magnetohydrodynamical turbulence this would imply that the inferred mean-field strength of the turbulent component
is $3.5 - 4.6$G. This value is reasonable for the magnetosphere of
a T Tauri star where typical surface fields are 0.3-1~kG
(Johns-Krull, 2007). However, this value would imply that the $\beta$ of the
atmospheric plasma is as high as  $0.14$!. 

To conclude, the observations presented in this letter indicate
that AK~Sco is the first pre-main sequence binary for which
evidence of a highly perturbed atmosphere has been reported.
However, conclusive evidence on the source of the perturbation
and the mechanism that drives it, cannot be inferred solely 
from these data. Further UV observations and monitorings are required.

\acknowledgments

This work has been partly financed by the Ministry of Education of
Spain through grant AYA2007-67726 and by the Comunidad Aut\'onoma
de Madrid by grant CAM-S-0505/ESP/0237.

 \facility{HST (STIS)}

\clearpage



\begin{figure}
\epsscale{.80}
\plotone{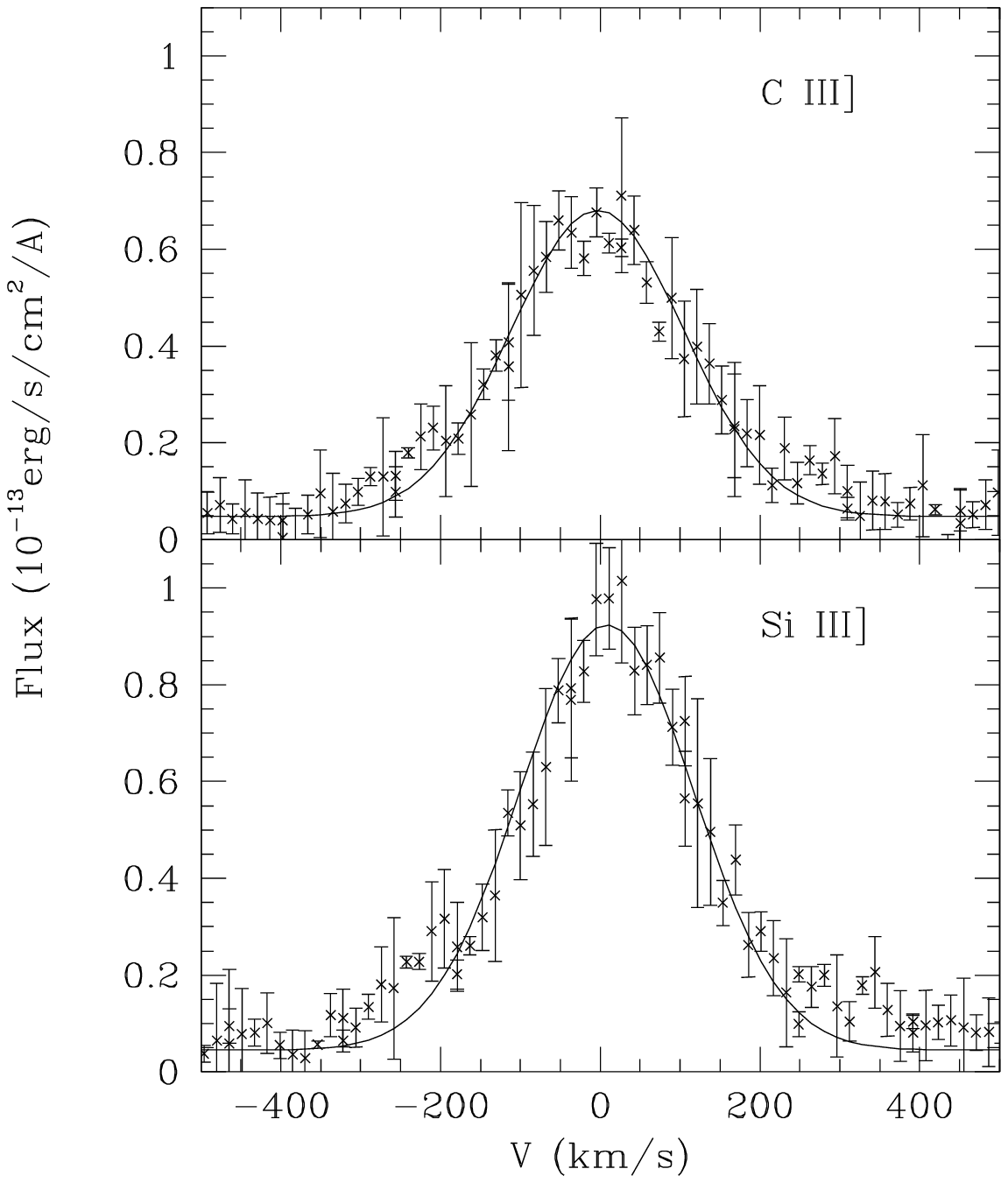}
\caption{C~III] (top) and Si~III] (bottom) profiles
of AK~Sco obtained with the {\it Hubble Space Telescope}
and the Space Telescope Imaging Spectrograph. Data have
been processed with the Routine Science Data Pipeline (RSDP).
The major source of  inaccuracy in the calibration is the centering
of the target in the aperture that can account for as much as
2.8~km~s$^{-1}$.  The exposure duration was 1944~s that were splitted
into three subexposures; the average value for each pixel and the deviation
from this average (error-bar) are plotted. A curve fit to
a single gaussian plus a constant is also shown. } \label{fig1}
\end{figure}

\clearpage

\begin{figure}
\plotone{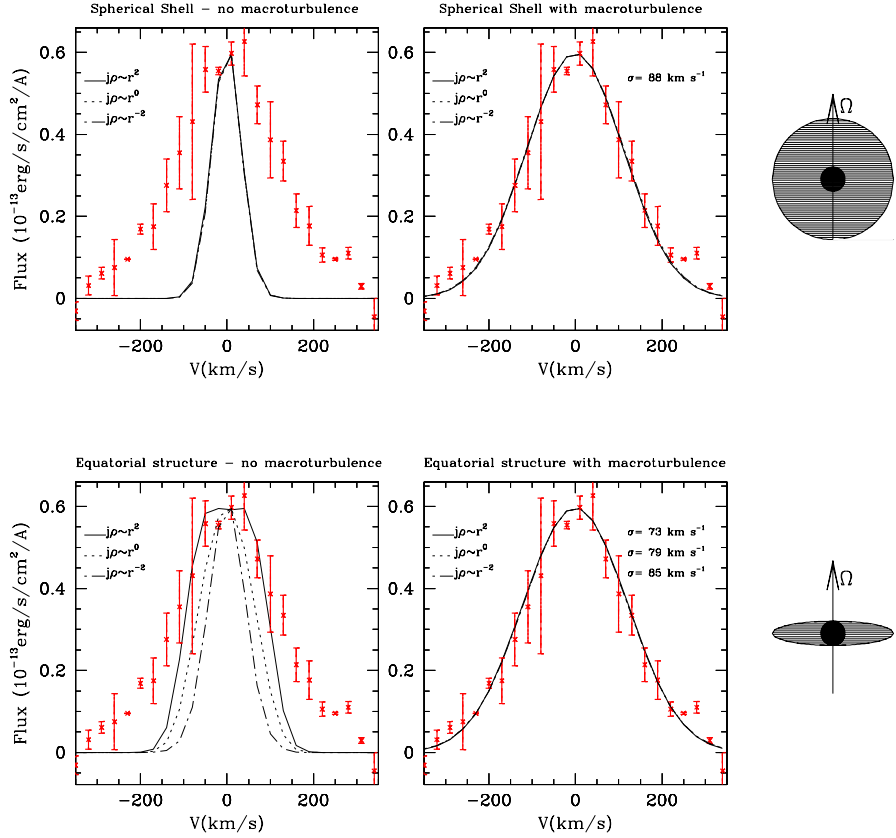} \caption{ Predicted C~III] line profiles for
equatorial and spherical distributions of the emitting plasma. Top
and bottom panels represent synthetic profiles for spherical-like
and disk-like mass distributions. Three types of dependences of
the line emissivity with the radius have been computed for the two
possible mass distributions: $\epsilon _{rad,l} \propto
r^{\alpha}$ with $\alpha = -2, 0$ and 2, as indicated in the
figures. The synthetic profiles are represented in the left
panels. It is necessary to add a macro turbulence field to fit the
profiles, as shown in the right panels.} \label{fig2}
\end{figure}

\clearpage

\begin{table}
\begin{center}
\caption{Main properties of AK Sco spectroscopic binary
system\tablenotemark{a}.\label{tbl-1}}
\begin{tabular}{ll}
\tableline\tableline
Property & Value \\
\tableline
Projected semimajor axis\tablenotemark{b}& $ a\sin i = 30.77 \pm 0.12$R$_{\odot}$ \\
Eccentricity\tablenotemark{b}\tablenotemark{c} & e= 0.47 \\
Orbital period\tablenotemark{b}\tablenotemark{c} & P=13.609 d \\
Inclination\tablenotemark{c} & $ i=65^o-70^o$ \\
Age\tablenotemark{c} & 10-30 Myrs  \\
Spectral type\tablenotemark{b}\tablenotemark{c}\tablenotemark{d} & F5\\
Stellar Mass\tablenotemark{c}\tablenotemark{d}  & $M_* = 1.35 \pm 0.07$M$_{\odot}$\\
Radius\tablenotemark{c}\tablenotemark{d} & $R_* = 1.59 \pm 0.35$R$_{\odot}$ \\
Projected rotation velocity\tablenotemark{b}\tablenotemark{c}\tablenotemark{d}
& $v \sin i = 18.5 \pm 1.0$ km s$^{-1}$ \\

\tableline
\end{tabular}
\tablenotetext{a}{The ephemeris is  $\tau (min) = HJD
2446666.380+13.609 \times E$ from Andersen et al. 1989}
\tablenotetext{b}{from Andersen et al. 1989}
\tablenotetext{c}{from Alencar et al. 2003}
\tablenotetext{d}{The
system is composed by two nearly identical stars so these
parameters are the same for both components.}
\end{center}
\end{table}

\end{document}